# A new network node similarity measure method and its applications

Yongli LI

School of Management, Harbin Institute of Technology, Harbin 150001, P.R.China;
E-mail: yongli.0440004@gmail.com

Peng LUO

School of Management, Harbin Institute of Technology, Harbin 150001, P.R.China;
E-mail: luopeng_hit@126.com

Chong WU

School of Management, Harbin Institute of Technology, Harbin 150001, P.R.China;
E-mail: wuc_hit@126.com

**ABSTRACT**

Network node similarity measure has been paid particular attention in the field of statistical physics. In this paper, we utilize the concept of information and information loss to measure the node similarity. The whole model is based on this idea that if two nodes are more similar than the others, then the information loss of seeing them as the same is less. The present new method has low algorithm complexity so that it can save much time and energy to deal with the large scale real- world network. We illustrate the availability of this approach based on two artificial examples and computer-generated networks by comparing its accuracy with the other selected approaches. The above tests demonstrate that the new method can provide more reasonable results consistent with our human common judgment. The new similarity measure method is also applied to predict the network evolution and predict unknown nodes' attributions in the two application examples.

**Acknowledgements:** This study was partly funded by National Natural Science Foundation of China (No.71271070) and China Scholarship Council (No.201306120159).

# 1. Introduction

Networks provide a powerful framework to model real world relationships in which nodes denote entities, and links indicate the interactions between these entities [1]. One problem from networks that attracted particular attention is that "network node similarity measure": the nodes in networks are often found to be similar to some others. It would be rather useful to be able to answer questions such as "how similar are these two vertices?" Of course, it is much sense to answer these questions. In the field of social network, for example, the nodes can represent people and the links can represent friendship between individuals [2,3]. If two people have similar interests, backgrounds, or friends, they may be similar to each other. Then, it can be economical and applicable to make friend recommendation [4-6], link prediction [7,8], peer-effect analysis [9,10], and so forth. Apartment from the above examples, a good and stylized method of network node similarity measure can also capture important aspects of a number of applications, such as in the field of physics [11,12], transport [13], biology [14], bibliometrics [15] and so forth. This paper aims to propose such a good and stylized method which would be not only a new method in the field of statistical physics but also a useful one covering large scopes of applications.

This paper will review the related work in the second part. Compared to these existing methods, ours has two distinctive features. First, our complementary approach is based on information theory [16,17]. This idea gives an information measure for a network and focuses on the information loss when two nodes are seen as the same. Of course, if the information loss of seeing two nodes as the same is lower than the other two, it means that the two nodes are more similar. It is noted that the basic principle of our model is contrast with the existing ones introduced in the related work of this paper, thus ours is innovative from this angle. Second, our method has the algorithm time complexity $O(n^2)$, where $n$ is the node number of a given network. The property facilities our method to be applied in the large network which is often faced in the real-world problem. In all, these two features together allow us to answer the above question with higher reasonability consistent with our human sense and lower algorithm complexity than majority of these existing methods.

To demonstrate the method's reasonability and present its calculation process clearly, we organize this paper as follows: in the following part 2, the related work is reviewed in brief; in part 3, the model and its algorithm is explained and showed in detail with mathematical proof and algorithm analysis; in part 4, an simple artificial example and a simulation-based example are given to uncover our method's properties and compare it with the selected existing methods; in part 5, two applications are given to illustrate the method's applicability, and the last part 6 concludes.

# 2. Related work

The most common approach adopted in previous work is to count how many neighbors two nodes have in common. It is in the sense that nodes are similar to the extent that their neighborhoods overlap. Let $N_i$ be the neighborhood of vertex $i$ in a network, i.e., the set of nodes that are directly connected to $i$ via an edge. Then the similarity measure for two nodes $i$ and $j$ is

$$\sigma_1(i,j) = |N_i \cap N_j|. \tag{1}$$

However, there is an drawback of this measure that the nodes with high degree are favored to be more similar than the low-degree nodes, because the high-degree vertices would have a large value even if only a small fraction of their neighbors are shared. Therefore, this method is not entirely satisfactory.

In one stream, there are many ways being proposed to overcome the above problem with time going by. One way is to normalize the number of shared nodes based on the size of its two neighborhoods' union.

$$\sigma_2(i,j) = \frac{|N_i \cap N_j|}{|N_i \cup N_j|},  \qquad (2)$$

which is commonly called the *Jaccard index* which was proposed in [18]. Then, the *cosine similarity* was proposed by Salton and was widely used in the literature on citation networks [19]. It is the cosine of the angle between the characteristic vectors of the two neighborhoods. We leave out the vertices *i* and *j* when counting the size of their neighborhoods, as this yields a better measure on loop-less graphs.

$$\sigma_3(i,j) = \frac{|N_i \cap N_j|}{\sqrt{|N_i||N_j|}}. \qquad (3)$$

Besides, there are also many another ways to improve the common similarity measure (1), like Ravasz et al. [20], Burt [21], and Goldberg and Roth [22] as shown below one by one:

$$\sigma_4(i,j) = \frac{|N_i \cap N_j|}{\min(|N_i|,|N_j|)}, \qquad (4)$$

$$\sigma_5(i,j) = \sqrt{|N_i \cap N_j|}, \qquad (5)$$

$$\sigma_6(i,j) = \left(|N_i \cap N_j|\right)^2. \qquad (6)$$

In the other stream, many researchers have been working to propose new node similarity measure methods in other ways. For example, Symeonidis et al. [23] defined a new way to calculate the similarity between vertices on the basis of the Tanomoto coefficient [24]. Firstly, they define the similarity measure as follows.

$$\sigma_7(i,j) = \frac{|N_i \cap N_j|}{|N_i| + |N_j| - |N_i \cap N_j|}. \qquad (7)$$

However, it is not reasonable enough since the similarity values between all non-neighbor nodes are zero based on formula (7). Then, they define a transitive node similarity which is calculated by the product of basic similarity between the nodes appearing in the shortest path. As a result, they get the following method.

$$\sigma_8(i,j) = \begin{cases} 0, \text{ if there is no path between the two nodes;} \\ \sigma_7(i,j), \text{ if } i,j \text{ are neighbors;} \\ \prod_{k=1}^{t} \sigma_7(v_k, v_{k+1}), \text{ otherwise.} \end{cases} \qquad (8)$$

where $v_1 = i, v_{t+1} = j$ and the nodes $v_k \, (k = 2, \cdots, t)$ are all the intermediate nodes that the shortest path from *i* to *j* passes through. They also use the transitive node similarity to predict link in social networks.

Recently, Chen et al. [25] introduced another new vertex similarity measure called *relation strength similarity* (RSS) which could better capture potential relationships of real world network structure. Firstly, they define the *relation strength*, a normalized edge weight score reflecting the relative degree of similarity between neighbor vertices. The relation strength between two nodes *i* and *j* can be calculated as follows.

$$\sigma_9(i,j) = \begin{cases} \dfrac{\alpha_{ij}}{\sum_{x \in n_i} \alpha_{ix}}, & \text{if } i \text{ and } j \text{ are adjacent,} \\ 0, & \text{otherwise.} \end{cases} \quad (9)$$

where $\alpha_{ij}$ is the weight of the edge between nodes $i$ and $j$, and $n_i$ is the set of neighbor vertices of node $i$. Secondly, if node $i$ and $j$ are not adjacent and they are linked by a path, the *generalized relation strength* is defined as

$$\sigma_{10}(i,j) = \begin{cases} \prod_{k=1}^{t} \sigma_9(v_k, v_{k+1}), & \text{if } t < r, \\ 0, & \text{otherwise.} \end{cases} \quad (10)$$

where $v_1 = i$, $v_{t+1} = j$ and the nodes $v_k \, (k = 2, \cdots, t)$ are all the intermediate nodes that form the path. The $r$ is a discovery range parameter to control the maximum degree of separation for a generalized relation strength calculation. For example, if assumed that there are $M$ simple paths from $i$ to $j$ with path length shorter than $r$, then the relation strength similarity between $i$ and $j$ is calculated by summing all the generalized relation strengths.

$$\sigma_{11}(i,j) = \sum_{m=1}^{M} \sigma_{10}(i,j). \quad (11)$$

Moreover, there are also many other researchers who make contributions to the node similarity. A stochastic approach for determining similarity of vertices was presented by Nowicki and Snijders [26]. Also, Leicht et al. [2] considered two nodes are similar if their neighbors in the network are similar. Accordingly they constructed a method for quantifying the similarity of nodes in networks based on this idea. Furthermore, Penner et al. [14] proposed a biologically motivated quantity, twinness, to evaluate local similarity of network nodes. Beside, Thiel and Berthold [27] took advantage of the spreading activation and gave two different kinds of node similarities measure methods.

### 3. Model and algorithm

The proposed model is established based on the idea that "If two nodes are more similar than the others, then the information loss of seeing them as the same is less than that of the others". In this section, we first present how to use a probability function to describe an undirected network, and then give the definition of "information loss" to measure the loss of seeing two different nodes as the same from the viewpoint of information theory.

Considering an undirected network **G** with $n$ nodes, its links can be described by the symmetric matrix $\mathbf{A} = [a_{ij}]_{n \times n}$, where

$$a_{ij} = \begin{cases} \text{weight between nodes } i \text{ and } j, & \text{when } i \neq j \\ \sum_{i=1}^{j-1} a_{ij} + \sum_{i=j+1}^{n} a_{ij}, & \text{when } i = j \end{cases},$$

Especially, as for an unweighted network,
when $i \neq j$,

$$a_{ij} = \begin{cases} 1, & \text{if there exists a link between nodes } i \text{ and } j \\ 0, & \text{otherwise} \end{cases};$$

when $i = j$,

$$a_{ij} = \sum_{i=1}^{j-1} a_{ij} + \sum_{i=j+1}^{n} a_{ij},$$

here, $a_{ij}$ is the degree of the node $i$.

Then, the joint probability density function of nodes $i$ and $j$ can be defined as

$$p(i,j) = \frac{a_{ij}}{\sum_{i=1}^{n}\sum_{j=1}^{n} a_{ij}}. \qquad (12)$$

The above given $p(i,j)$ is the description of an undirected network from the angle of probability function. Also, the $p(i,j)$ can be checked to satisfy the definition of a probability function. Besides, we have $p(i) = \sum_{j=1}^{n} p(i,j)$, and so is the $p(j)$.

The definition of "Mutual information" in the field of information theory can be applied to measure how much information is contained in a complex network. The information of a given network with $n$ nodes denoted by $I(\mathbf{N};\mathbf{N})$ is defined as

$$I(\mathbf{N};\mathbf{N}) = \sum_{i \in \mathbf{N}} \sum_{j \in \mathbf{N}} p(i,j) \cdot \log \frac{p(i,j)}{p(i) \cdot p(j)}, \qquad (13)$$

where, $\mathbf{N}$ is the set of nodes. When two nodes (for example $n_1$ and $n_2$) are seen as the same, the new node set donated by $\mathbf{N}^*$ is

$$\mathbf{N}^* = \mathbf{N} - \{n_1\} - \{n_2\} + \{<n_1, n_2>\}, \qquad (14)$$

which has $n-1$ nodes. Then, the mutual information $I(\mathbf{N};\mathbf{N}^*)$ between the original node set ($\mathbf{N}$) and the new node set ($\mathbf{N}^*$) is

$$\begin{aligned}I(\mathbf{N};\mathbf{N}^*) &= \sum_{i \in \mathbf{N}} \sum_{j \in \mathbf{N}^*} p(i,j) \cdot \log \frac{p(i,j)}{p(i) \cdot p(j)} \\ &= I(\mathbf{N};\mathbf{N}) - \sum_{i \in \mathbf{N}} \sum_{k=1}^{2} p(i,n_k) \cdot \log \frac{p(i,n_k)}{p(i) \cdot p(n_k)} + \sum_{i \in \mathbf{N}} p(i,<n_1,n_2>) \cdot \log \frac{p(i,<n_1,n_2>)}{p(i) \cdot p(<n_1,n_2>)}.\end{aligned}$$
(15)

The difference between $I(\mathbf{N};\mathbf{N}^*)$ and $I(\mathbf{N};\mathbf{N}^*)$ uncovers the information loss when the two nodes $n_1$ and $n_2$ are seen as the same. That is to say, the information loss is the cost of taking the two nodes undifferentiated, which means the less the information loss is, the more close or the more similar the two nodes are. In detail, the **information loss** denoted by $\Delta I(n_1, n_2)$ is

$$\begin{aligned}\Delta I(n_1,n_2) &= I(\mathbf{N};\mathbf{N}) - I(\mathbf{N};\mathbf{N}^*) \\ &= \sum_{i \in \mathbf{N}} \sum_{k=1}^{2} p(i,n_k) \cdot \log \frac{p(i,n_k)}{p(i) \cdot p(n_k)} - \sum_{i \in \mathbf{N}} p(i,<n_1,n_2>) \cdot \log \frac{p(i,<n_1,n_2>)}{p(i) \cdot p(<n_1,n_2>)}. \end{aligned} \qquad (16)$$

Because the two nodes $n_1$ and $n_2$ are seen as the same, namely they are merged into one node, it holds

$$p(i,<n_1,n_2>) = p(i,n_1) + p(i,n_2), \qquad (17)$$
$$p(<n_1,n_2>) = p(n_1) + p(n_2), \qquad (18)$$
$$p(i|<n_1,n_2>) = \frac{p(i,n_1) + p(i,n_2)}{p(n_1) + p(n_2)}. \qquad (19)$$

Accordingly, $\Delta I(n_1, n_2)$ can be calculated to be the following form, where it is noted that when $p(k|n_k) = 0$, we let $p(k|n_k) \cdot \log p(k|n_k) = 0$ $(k=1,2)$.

$$\Delta I(n_1, n_2) = p(n_1) \cdot \sum_{k \in \mathbf{N}} p(k \mid n_1) \cdot \log \frac{p(k \mid n_1)}{p(k \mid <n_1, n_2>)} + p(n_2) \cdot \sum_{k \in \mathbf{N}} p(k \mid n_2) \cdot \log \frac{p(k \mid n_2)}{p(k \mid <n_1, n_2>)}.$$
(20)

Besides, $\Delta I(n_1, n_2)$ must be no less than zero which is given and proved in Property 1. The property is consist with the nature rule that the information loss of regarding two different nodes as the same should be above or at least equal to 0.

[**Property 1**] $\Delta I(n_1, n_2) \geq 0$.

***Proof***. Based on that $x \log x$ is the convex function, when $\beta_1 + \beta_2 = 1$, it holds that
$$\beta_1 \cdot a \log a + \beta_2 \cdot b \log b \geq (\beta_1 a + \beta_2 b) \cdot \log(\beta_1 a + \beta_2 b),$$
(21)

When $p(n_1) + p(n_2) \neq 0$, let $\beta_1 = p(n_1)/(p(n_1) + p(n_2))$, $\beta_2 = p(n_2)/(p(n_1) + p(n_2))$, $a = p(k \mid n_1)$ and $b = p(k \mid n_2)$. As a result, for any $k \in \mathbf{N}$, the above inequality is

$$\frac{p(n_1)}{p(n_1) + p(n_2)} \cdot p(k \mid n_1) \cdot \log \frac{p(k \mid n_1)}{p(k \mid <n_1, n_2>)} + \frac{p(n_2)}{p(n_1) + p(n_2)} \cdot p(k \mid n_2) \cdot \log \frac{p(k \mid n_2)}{p(k \mid <n_1, n_2>)} \geq 0,$$

which means $\Delta I(n_1, n_2) \geq 0$ when summing the above formula based on all the $k \in \mathbf{N}$.

When $p(n_1) + p(n_2) = 0$, we have $p(n_1) = 0$ and $p(n_2) = 0$. Then, the two nodes $n_1$ and $n_2$ are isolated nodes which are not necessary to use the above model to analyze. Even so, in this case, it holds that $\Delta I(n_1, n_2) = 0$. □

In order to present the whole calculating process clear, we summarize the corresponding algorithm in the following Table 1. The total time complexity can be estimated as $O(n^2)$ from the following algorithm, where *n* is the number of nodes in the given network **G**. While, the time complexities of Symeonidis' method [23] and Chen's method [25] are both $O(n^3)$ which is higher than our method. From this viewpoint, our method is more suitable for large network than the just mentioned two methods.

Table 1　Algorithm

**Input**: the given undirected network **G** (*n* nodes and its relationship matrix **A**);
**Output**: the information loss of any two nodes in the given network **G**;
**Initialization**:
  (1) delete the isolated nodes;
  (2) calculate $p(i, j)$ of the given network **G**;
  (3) calculate $p(i)$ and $p(j)$;
  (4) $a \leftarrow 0$ and $b \leftarrow 0$.
**Loop**:
  for  *i*=1 to *n*
    for  *j*=*i*+1 to *n*
      for  *k*=1 to *n*
        $p(k \mid i) = p(k, i) / p(i)$;
        $p(k \mid j) = p(k, j) / p(j)$;
        $p(k \mid <i, j>) = (p(k, i) + p(k, j)) / (p(i) + p(j))$;
        if  $p(k \mid i) \neq 0$
          then  $a \leftarrow a + p(k \mid i) \cdot (\log p(k \mid i) - \log p(k \mid <i, j>))$;
        else
          CONTINUE;

```
                    end
                if  p(k|n_2) ≠ 0
                    then  b ← b + p(k|j)·(log p(k|j) − log p(k|<i,j>));
                else
                    CONTINUE;
                end
            end
        ΔI(i,j) = p(i)·a + p(j)·b;
        end
    end
    print: the information loss  ΔI(i,j)  of any two nodes  (i,j).
```
―――――――――――――――――――――――――――――――――――――――――

To make the expression clear, we further define $1/\Delta I(i,j)$ as the node similarity measure between two nodes *i* and *j* based on the above **information loss** proposed in this paper. The definition is consistent with the common definition of network node similarity measure, for which the larger value means higher similarity between the corresponding two nodes.

## 4. Examples
### 4.1. Simple artificial examples

Our method can be used to both networks with or without weights. However, except the Chen's method which can be used to both kinds, the other methods introduced in the part of related work can only be applicable in the case of networks without weight. Here, we design two artificial networks, and compare our method (OUM for short) with cosine similarity method (CSM) shown by formula (3), Symeonidis's method (SYM) shown by formulas (7) to (8) and the Chen's method (CHM) showed by formulas (9) to (11).

The two simple artificial networks are shown in Fig.1 and Fig.2 respectively, where the weights in Fig.2 reflect relation forces between two nodes. The so-called relation force has many manifestations in real life, for example, the number of common hobbies between friends, the times of collaborations between authors, the degree of finance flows between companies, and so forth.

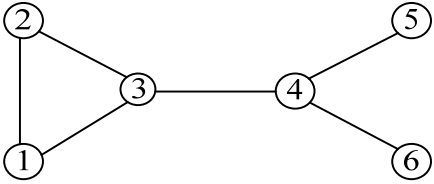
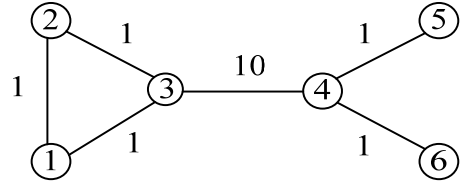

Fig.1 The artificial network without weight    Fig.2 The artificial network with weights

Intuitively, the node similarity measure is not only dependent on the nodes' positions, but also the relation forces between nodes. Accordingly, from the above two figures, we can judge, in our sense, that node 1 and node 2 are the most similar nodes in Fig.1 and the similarity measure value between node 3 and node 4 in Fig.2 should be higher than that in Fig.1. Besides, according to the weight between nodes 3 and 4 becoming larger compared to that in Fig.1, the left part and the right part in Fig.2 will be much closer than the two parts in Fig.1. As a result, similarity measure values of nodes pairs (1,5), (1,6), (2,5) and (2,6) in Fig.2 should be larger than their counterparts in Fig.1. If a similarity measure method is a good one, it should be consistent with our judgment. Let us compare the similarity measure values from four methods.

Table 2   The simple artificial examples' results

| Figure | Fig.1 | | | | Fig.2 | |
| --- | --- | --- | --- | --- | --- | --- |
| node pair | OUM | CSM | SYM | CHM | OUM | CHM |
| **(1,2)** | **70.6315** | **1.000** | **0.3333** | 0.6667 | **176.5848** | 0.5417 |
| (1,3) | 20.2544 | 0.7071 | 0.2500 | **0.7500** | 12.8702 | 0.7500 |
| (1,4) | 4.4912 | 0.4082 | 0.0000 | 0.2500 | 7.3780 | 0.6250 |
| **(1,5)** | 6.2842 | 0.0000 | 0.0000 | 0.0833 | 15.7107 | 0.0521 |
| **(1,6)** | 6.2842 | 0.0000 | 0.0000 | 0.0833 | 15.7107 | 0.0521 |
| (2,3) | 20.2544 | 0.7071 | 0.2500 | **0.7500** | 12.8702 | 0.7500 |
| (2,4) | 4.4912 | 0.4082 | 0.0000 | 0.2500 | 7.3780 | 0.6250 |
| **(2,5)** | 6.2842 | 0.0000 | 0.0000 | 0.0833 | 15.7107 | 0.0521 |
| **(2,6)** | 6.2842 | 0.0000 | 0.0000 | 0.0833 | 15.7107 | 0.0521 |
| **(3,4)** | 6.2842 | 0.0000 | 0.0000 | 0.3333 | 20.3070 | **0.8333** |
| (3,5) | 7.7111 | 0.0000 | 0.0000 | 0.1111 | 16.2167 | 0.0694 |
| (3,6) | 7.7111 | 0.0000 | 0.0000 | 0.1111 | 16.2167 | 0.0694 |
| (4,5) | 27.8087 | 0.0000 | 0.0000 | 0.3333 | 28.0481 | 0.0833 |
| (4,6) | 27.8087 | 0.0000 | 0.0000 | 0.3333 | 28.0481 | 0.0833 |
| (5,6) | 17.3124 | 0.0000 | 0.0000 | 0.3333 | 43.2807 | 0.0833 |

The above results demonstrate that only the result of our method completely agrees with the three pieces of judgments. First, CSM and SYM can only deal with the unweighted case, and their results are not elaborate, or in another word, undistinguished, since many node pairs' similarity measure values are zero. Second, the result of CHM violated the first and the third pieces of judgment, which means that its result may not be accordant with our most human common sense. Thus, from this given simple artificial example, our method is prior to the other three to some extent.

**4.2. Comparisons based on simulation analysis**

We continue to rest our method on a series of networks generated by computers. Two things are important for the simulation analysis: one is how to generate the computer-based networks; the other is how to choose the index for comparing these results of different methods.

As for generating the computer-based networks, we consider the *Community Structure* to be the benchmark for deciding which nodes are similar and which ones are not. That is to say, if two nodes in one community, we regard them similar, and vise versa. Based on this idea, we follow Girvan and Newman's way [28] to produce randomly artificial networks that have known community structure, which has been a famous and widely used simulation method for testing [29]. Each generated graph is constructed with 128 nodes divided into 4 communities of 32 nodes each. Edges are placed between node pairs independently at random, with probability $P_{in}$ for nodes belonging to the same community and $P_{out}$ for nodes in different communities. When the $P_{in}$ is given, the $P_{out}$ can be calculated out to keep the average degree of nodes as 28, since keeping the same average degree of nodes can make different networks comparable. Then, we denote the computer-generated graphs as $RN(4,32,28,P_{in})$ where $P_{in}$ varies from 0.60 to 0.95 by 0.05 in each step, and we can get 100 artificial networks for each $P_{in}$ and average the results to make the comparison analysis with SYM and CHM mentioned above.

As for the index, we have known nodes within one community similar to each other, thus each node has 31 similar nodes in the network. We can rank all these other nodes based on their similarity measure with the given one, and compare the first 31 nodes with our known similar nodes of the given one. Let us denote the number of wrong similar nodes as to node *i* as $N_i$, then we can get the index *N* by summing all these $N_i$ according to *i*. Of course, the smaller the *N* is, the better the corresponding result is. Accordingly, the **error rate** can be defined as $N/(128\times31)$. Based on these 100 generated artificial networks for each $P_{in}$, we can get a comprehensive index value by averaging and its corresponding standard deviation. The whole result is shown in the following Fig.3.

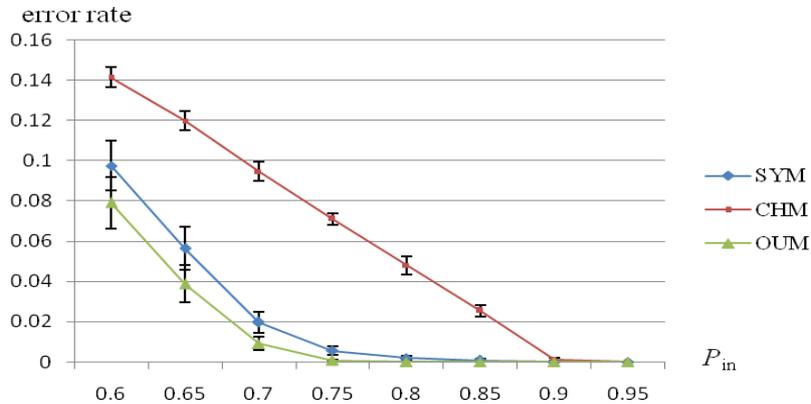

Fig.3  The results based on computer-generated networks

The Fig.3 illustrates that when $P_{in} \geq 0.75$, our method performs nearly perfectly with classifying nearly 100% nodes correctly. With $P_{in}$ varying from 0.6 to 0.75, the error rate is below 0.08 which is much less than those of CHM and SYM. The result indicates that our approach is much more efficient in terms of finding the similarity nodes. Based on the above tests and comparisons, we can conclude that our approach is efficient in searching similar node, and in these experiments, it performs better than the listed approaches in terms of accuracy. Although these artificial networks in the above experiments cannot represent all the real world networks, they can prove the availability and superiority of our approach to some extent.

## 5. Applications
In the previous section we tested our method on several artificial networks for which the node similarity was given beforehand. The results indicate that our method is a sensitive and accurate method for extracting node similarity. In this section, we apply our method to two real networks for which the node similarity was not directly given but some node attributions are given. In these cases, we infer which nodes are similar according to their attributions and then check the results of our method with our inference. The application can help us to further test our method and to understand the reasons and origins of node similarity in the complex, real-world and tangled datasets. Our first example is a friendship network from Zachary's Karate club; our second is a collaboration network of scientists.

### 5.1. Friendship network
The first application here is drawn from the well known Zachary's Karate club [30]. The 34 members in this club forms a friendship network which is a weighted network. In Zachary's study, he observed the 34 members over two years and found a disagreement between the administrator of the club and the club's instructor. As a result, the instructor left and started a new club, taking some of the club's members away. One interesting thing is whether the friendship network of these 34 members takes effect in the process of forming the new club. In this paper, we will study this real-world problem from the viewpoint of node similarity. First of all, we draw Fig.4 to show the friendship network and the split two clubs.

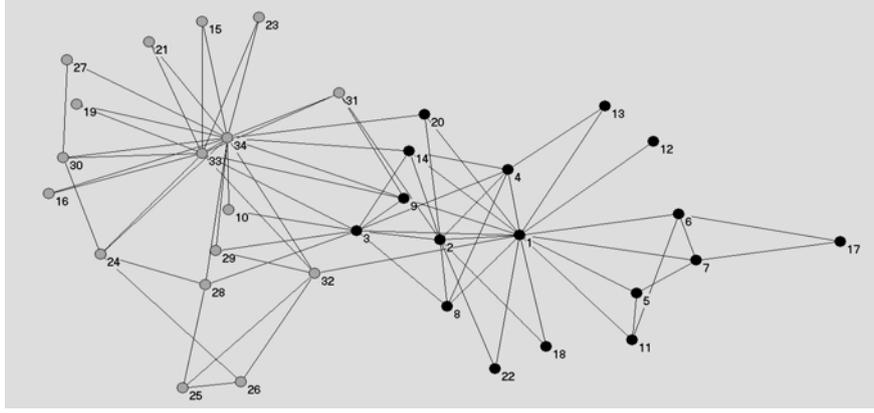

Fig.4  The friendship network from Zachary's Karate club

Then, our method is applied to measure the similarity between any two members, namely any two nodes in the network. We infer that the two nodes with large similarity measure are much likely to get together when the two clubs form. Thus, like the part 4.2, we can test whether the first four similar nodes of each node belong to the same club with the tested node. The whole results are shown in Table 3, where $N_i$ means the number of the node $i$'s similar nodes not within the same club with it. From this table, we can find that most of nodes and their first four similar nodes are in the same community. The total percentage of the first four similar nodes that don't belong to the same club with the tested node is below 10%. We can use our method to predict which nodes get together into the same club when the two clubs form. Thus, in this application, our new node similarity measure performs well in predicting the network evolution.

Table 3 Results of the first application

| Node | The first four similar nodes | $N_i$ | Node | The first four similar nodes | $N_i$ |
|---|---|---|---|---|---|
| 01 | 12, 18, 22, 20 | 0 | 18 | 22, 12, 20, 13 | 0 |
| 02 | 22, 18, 20, 12 | 0 | 19 | 23, 21, 10, 15 | 0 |
| 03 | 18, 10, 12, 22 | 1 | 20 | 18, 22, 12, 10 | 1 |
| 04 | 13, 18, 08, 12 | 0 | 21 | 19, 15, 23, 10 | 0 |
| 05 | 11, 12, 18, 22 | 0 | 22 | 18, 12, 20, 13 | 0 |
| 06 | 17, 07, 11, 12 | 0 | 23 | 19, 21, 15, 10 | 0 |
| 07 | 17, 06, 12, 18 | 0 | 24 | 19, 21, 10, 15 | 0 |
| 08 | 18, 22, 12, 13 | 0 | 25 | 26, 28, 10, 19 | 0 |
| 09 | 10, 19, 31, 12 | 3 | 26 | 25, 32, 10, 19 | 0 |
| 10 | 19, 29, 23, 21 | 0 | 27 | 30, 10, 19, 12 | 1 |
| 11 | 05, 12, 18, 06 | 0 | 28 | 10, 19, 25, 12 | 1 |
| 12 | 18, 22, 01, 20 | 0 | 29 | 10, 19, 12, 18 | 2 |
| 13 | 04, 12, 18, 22 | 0 | 30 | 27, 19, 21, 10 | 0 |
| 14 | 18, 10, 12, 22 | 1 | 31 | 19, 21, 10, 15 | 0 |
| 15 | 19, 21, 23, 10 | 0 | 32 | 19, 29, 10, 21 | 0 |
| 16 | 19, 21, 23, 15 | 0 | 33 | 21, 19, 15, 23 | 0 |
| 17 | 07, 06, 12, 18 | 0 | 34 | 19, 10, 23, 15 | 0 |

### 5.2. Collaboration network

The second application is based on a collaboration network of scientists [31]. There are 1589 nodes in this network which represent researchers from a broad variety of fields. An edge is formed between a pair of researchers if they coauthored at least one paper. From the statistical analysis, every scientist has approximately four authors on average. Besides, the main research subject of every researcher is also included in this dataset. Here, we want to use our method to predict the research subject of one researcher, namely, given one researcher's research subject, we use our method and the collaboration network to find his or her similar node in order to predict the similar node's research subject based on the given information. To test our method, we can check the predicted result with the fact since everyone's research subject is given in this network. If this method goes well in this application, we can focus on studying only several nodes in one network and then infer that its similar nodes would have the same attributions with our known one based on the responding formed network and our new method. First of all, the collaboration network is shown in Fig.5 with distinguishing them according to their research subjects.

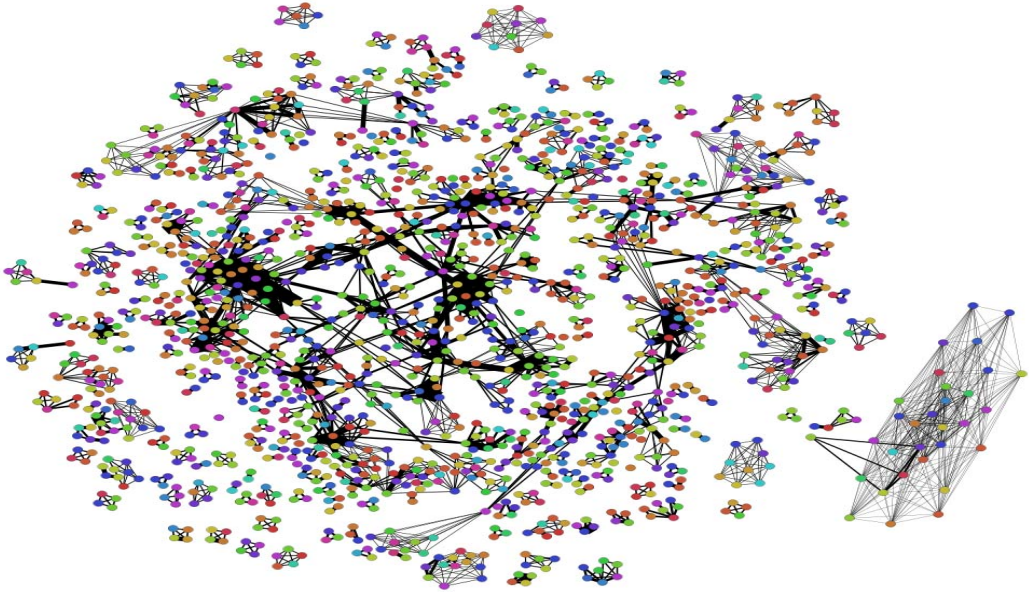

Fig.5    Structure graph of the collaboration network

Next, we randomly select one node and apply the proposed method to find its two most similar nodes. We infer the similar nodes' research subject using the given node's information, and compare it with the fact one. After repeating it 500 times, the total accuracy is approximately 83%. From this result, our method thus can infer the attributions of the similar nodes precisely based on several nodes that we have studied. Accordingly, we only need to research a few nodes when facing a new network; then, the other nodes' attributes can be inferred by their similar nodes. It can be a time-saving and labor-saving way to study a new network, especially a large-scale network.

**6. Conclusions and future work**

This paper proposes a new method to measure network node similarity. Contrast to the existing methods, the new method has the following features: (1) The new method utilizes the concept of information and information loss to measure the node similarity. Based on this concept, we point out that if two nodes are more similar than the others, then the information loss of seeing them as the same is also less. Accordingly, several mathematical formulas are deduced based on information theory. (2) The new method has the algorithm complexity $O(n^2)$, for which some large real-world networks can be solved with low time cost. (3) Illustrated from the artificial examples and applications, the new method can give more reasonable results which are more consistent with our common judgment.

The new method can be further extended in theory and in practice, which could be the future work. In theory, the method can be the first step to detect the community structure in one network. Since the concept of information loss can be used to find the similar nodes, the concept can also be used to measure the quantitative loss of getting nodes together into one community. Following this idea, one new method for exploring the communities of networks can be created. In practice, we have given some new ideas for using the new method in the application part of this paper. Especially, finding the similar nodes can predict the network evolution as the first application shows and also can predict the other unknown nodes' attributions as the second application shows. Thus, applying this method properly can not only guide the decisions but also save the time and energy. There are also large space left for using the new method in many fields such as goods recommendation, link prediction, epidemic control and so forth. The last not the least, we deeply hope the idea and the method presented here will prove applicable and useful in the analysis of many other networks.